\documentclass[a4paper,twocolumn,11pt,accepted=2026-09-11]{quantumarticle}
\pdfoutput=1
\usepackage[utf8]{inputenc}
\usepackage[english]{babel}
\usepackage[T1]{fontenc}
\usepackage{amsmath}
\usepackage{hyperref}
\usepackage[numbers,sort&compress]{natbib}

\usepackage{graphicx}
\usepackage{amssymb}
\usepackage{amsthm}
\usepackage{bbold}
\usepackage{pifont}
\usepackage{array}
\usepackage{tabularx}
\usepackage{xcolor}
\usepackage{psfrag}
\usepackage{soul}
\usepackage{textcomp}
\usepackage{bm}
\usepackage{bbm}
\usepackage{mathtools}

\usepackage{lipsum}
\usepackage{dcolumn}
\usepackage{tikz}
\usetikzlibrary{quantikz2}
\let\oldpercent\%
\renewcommand{\%}{\scalebox{0.85}{\oldpercent}}

\usepackage{tikz}
\usepackage{lipsum}

\begin{document}

\title{Enhanced measurements on quantum computers via the simultaneous probing of non-commuting Pauli operators}

\author{Rick P. A. Simon}
\affiliation{Department of Physics and Astronomy, University of Exeter, Stocker Road, Exeter EX4 4QL, UK}
\email{r.p.a.simon@exeter.ac.uk}
\orcid{0009-0009-3111-2615}

\author{Zheng Shi}
\email{zheng.shi@uwaterloo.ca}
\affiliation{Institute for Quantum Computing, University of Waterloo, Waterloo, ON N2L 3G1, Canada}

\author{Charlie Nation}
\email{c.nation2@exeter.ac.uk}
\affiliation{Department of Physics and Astronomy, University of Exeter, Stocker Road, Exeter EX4 4QL, UK}

\author{Andrew Jena}
\email{ajjena@uwaterloo.ca}
\affiliation{Institute for Quantum Computing, University of Waterloo, Waterloo, ON N2L 3G1, Canada}
\affiliation{Department of Combinatorics \& Optimization, University of Waterloo, Waterloo, ON N2L 3G1, Canada}

\author{Luca Dellantonio}
\email{l.dellantonio@exeter.ac.uk}
\affiliation{Department of Physics and Astronomy, University of Exeter, Stocker Road, Exeter EX4 4QL, UK}

\maketitle

\begin{abstract}
  Measuring the state of quantum computers is a highly non-trivial task, with implications for virtually all quantum algorithms. A promising avenue is multi-copy schemes, where identical copies of a quantum state are measured jointly so that all Pauli operators within the considered observable can be simultaneously assessed. Here, we present a first implementation of such a two-copy scheme in a measurement protocol. Based on Bayesian statistics, it accurately estimates not only the average of the desired observable but also the error en route. This enables an adaptive shot-allocation algorithm that preferentially samples the most uncertain Pauli terms. In regimes with many non-commuting Pauli operators, this ``double'' scheme can outperform the state-of-the-art measurement protocol in minimizing total shots for a given precision. We also numerically confirm the finding in previous theoretical works that the two-copy scheme incurs an overhead due to the square-root relationship between the variance of measured quantities and the number of measurement shots.
\end{abstract}

\section{Introduction}
\label{sec:intro}

An essential task for all quantum algorithms is to accurately estimate the expectation value of an observable \cite{Lloyd1996,Georgescu2014,Daley2022}, expressed in terms of Pauli strings (PS), within a given measurement budget $M$. Groups of commuting PS can be measured simultaneously to lower the required number of measurements. Nevertheless, to attain the desired estimation accuracy, in practice one often needs to repeatedly probe many such groups of commuting PS. This requires an efficient strategy to allocate measurement budgets to different groups \cite{shlosberg2023adaptive,rick_tbp,Izmaylov2019,McClean2016,Wu2023,Jena2019,Gokhale2019,Verteletskyi2020,Yen2020,Hamamura2020,Crawford2021,Hadfield2021,Hadfield2022,Cai2025,Hu2025,Aaronson2017,Huang2021,Huang2020,chen2021exponential,Huang2022quantum,Chen2024optimal,King2024triply,Kohda2022,Arrasmith2020,Hillmich2021,Markovich2025quantum,Torlai2020}. 

Various such schemes have been developed. One class of methods groups PS into commuting subsets to measure them simultaneously \cite{McClean2016,Izmaylov2019,Jena2019,Gokhale2019,Verteletskyi2020,Hamamura2020,Yen2020,Crawford2021,Wu2023,shlosberg2023adaptive,rick_tbp}. Another approach, broadly called classical shadows \cite{Aaronson2017,Huang2020, Huang2021,Hadfield2022,Huang2022quantum,Hadfield2021,Cai2025,Chen2024optimal,King2024triply,Hu2025,chen2021exponential}, uses randomized measurements to build a classical representation of the state from which observables can be estimated. These techniques have greatly enhanced the estimation accuracies of observables measured on quantum computers. 
However, many successful approaches such as the variational ones \cite{McClean2016,Cerezo2021,Ferguson2021} need better measurement protocols. These variational algorithms prepare a parameterized quantum state and measure the constituent Pauli strings of the observable to learn a cost function, which is minimised by classical optimizers that iteratively update the state parameters. Tackling larger and larger systems \cite{Chakrabarti2021,Kitai2020,Cao2018,Paulson2021,Maurizio2025quantum} restricts the number of available cost function evaluations, and thus jeopardizes the future of these algorithms unless efficient measurement protocols are employed \cite{Barligea2025scalability}.

\begin{figure*}[ht!]
    \centering
    \includegraphics[width=\textwidth]{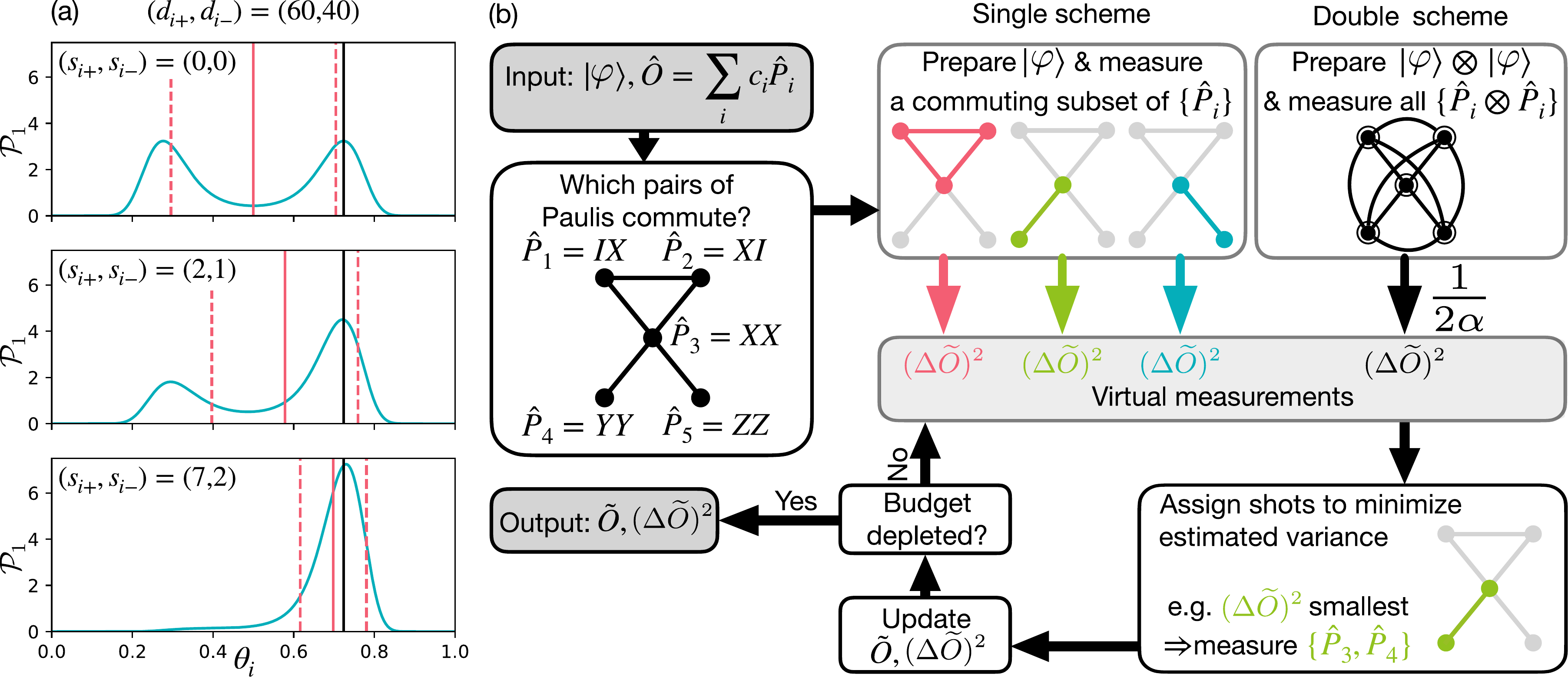}
    \caption{
    (a) The posterior $\mathcal{P}_1$ from Eq.~\eqref{eq:Bayes_Estimation_post} (blue line). The sign uncertainty is evident from the two peaks, which rapidly become one when shots are allocated without the double scheme (increase of $s_{i\pm}$, top to bottom). The vertical black line is the value $\theta_i = 0.725$ [$\implies \phi_i = 0.60125$, see Eq.~\eqref{eq:phi_i}] used to randomly generate the measurement outcomes $s_{i\pm}$ and $d_{i\pm}$. The pink vertical lines are the averages $\widetilde{\theta}_i$ (full) and the confidence interval $\widetilde{\theta}_i \pm \sqrt{(\Delta \widetilde{O})^2}$ (dashed), calculated from Eq.~\eqref{eq:Bayes_Estimation_exp}.
    (b) Scheme of our algorithm in Sec.~\ref{sec:Algo}, for the toy example $\hat{O}=\hat{IX} + \hat{XI} + \hat{XX} + \hat{YY} + \hat{ZZ}$
    ($c_i = 1$ for all $i=1,\dots,5$). These PS are depicted as vertices of a graph, connected when they commute \cite{shlosberg2023adaptive, rick_tbp}. All-to-all connected groups (pink, green, and blue) can be simultaneously measured. Alternatively, one can employ the more expensive (factor $1/(2\alpha)$, see main text) double scheme to assess the magnitude of all $\hat{P}_{i}$. To choose which group to probe, we assign a virtual measurement [Sec.~\ref{sec:Algo} and Eqs.~\eqref{eq:virtual_msmnt}] that predicts the group that minimizes $(\Delta \widetilde{O})^2$ (based on previous measurements, green in the figure as an example). After the real measurement, $\widetilde{O}$ and $(\Delta \widetilde{O})^2$ are updated following the methods in Sec.~\ref{sec:Theory}, and are either outputted by the algorithm (if the budget $M$ is depleted) or employed to assign the next measurement.  
    }
    \label{fig:schematics}
\end{figure*}

Of the strategies proposed so far, multi-copy schemes \cite{Huang2022quantum,chen2021exponential,Chen2024optimal,King2024triply} take advantage of the following observation. Two PS $\hat{P}_i$ and $\hat{P}_j$ composed of qubit operators (Pauli matrices, see Sec.~\ref{sec:Theory}) either commute or anti-commute with each other. Therefore, their ``double'' versions -- the tensor product of each PS with itself, $\hat{P}_i \otimes \hat{P}_i$ and $\hat{P}_j \otimes \hat{P}_j$ -- always commute and can therefore be measured simultaneously, the advantage of which is illustrated in Fig.~\ref{fig:schematics}(a). In this work, we implement the theoretical idea in Refs.~\cite{Huang2022quantum,chen2021exponential,Chen2024optimal,King2024triply} in a ``double'' scheme, which combines the multi-copy principle with a Bayesian approach to accurately estimate both the mean \emph{and} the statistical error of an arbitrary multi-qubit observable.

A major step forward compared to Refs.~\cite{Huang2022quantum,chen2021exponential,Chen2024optimal,King2024triply} is the ability to monitor the estimated error, which allows our scheme to predict how to allocate each measurement shot (either a group of commuting PS $\hat{P}_i$ or all of their double $\hat{P}_i \otimes \hat{P}_i$) so that the resulting error is minimized. We show numerically that when many PS do not commute with each other (so grouping is hard), our double scheme outperforms state-of-the-art measurement protocols, particularly for a limited measurement budget $M$. A paradigmatic class of observables that are characterized by long-range interactions between qubits and therefore includes several non-commuting PS is molecular Hamiltonians. These are a prime target for both near- and long-term quantum simulations \cite{Cao2018,Cao2019,Motta2021,Mishmash2023,Richerme2023}.

\section{Theory}
\label{sec:Theory}

The aim of this work is to build an accurate estimator $\widetilde{O} \approx \langle\hat{O}\rangle$, where $\hat{O} = \sum_{i=1}^{p} c_i \hat{P}_i$ is an observable defined on $q$ qubits and expressed as a weighted sum of $p$ PS $\hat{P}_i = \bigotimes_{j=1}^{q} \hat{\sigma}^{(i)}_{j}$. 
Here, $c_i$ are real constants and the Pauli operator $\hat{\sigma}^{(i)}_{j} \in \{\hat{I}_j,\hat{X}_j,\hat{Y}_j,\hat{Z}_j\}$ acts on qubit $j$.
Calling $\theta_i$ the probability of obtaining $+1$ when measuring $\hat{P}_i$ we have $\widetilde{O} = \sum_{i=1}^{p} c_i (2\widetilde{\theta}_i - 1)$. 
As explained in Sec.~\ref{sec:Algo}, we achieve high accuracy in $\widetilde{O}$ by continuously monitoring its estimated error $(\Delta\widetilde{O})^2$. Differently from elsewhere \cite{Dudley1978central,shlosberg2023adaptive,rick_tbp}, where $(\Delta\widetilde{O})^2$ is calculated from the variances and covariances of the PS, in this work $(\Delta\widetilde{O})^2$ is the variance of the estimate of $\langle\hat{O}\rangle$. 
The quantities $\widetilde{O}$ and $(\Delta\widetilde{O})^2$, central in this work, are thus determined via 
\begin{subequations} \label{eq:Estimations}
\small
    \begin{align}
        \widetilde{O} 
        &= 
        \sum_{i=1}^{p} c_i \widetilde{P}_i 
        = 
        \sum_{i=1}^{p} c_i (2\widetilde{\theta}_i - 1)
        \\
        (\Delta\widetilde{O})^2 
        &= 
        \widetilde{\langle \hat{O}\rangle^2} - \left(\widetilde{O}\right)^2 
        \label{eq:Estimations_Err} \\
        &=
        4\sum_i c_i^2(\widetilde{\theta^2_i} - \widetilde{\theta}_i^2) + 4 \sum_{i\neq j} c_i c_j (\widetilde{\theta_i\theta_j} - \widetilde{\theta}_i \widetilde{\theta}_j)
        , \nonumber
    \end{align}
\end{subequations}
where we highlight that $(\Delta\widetilde{O})^2$ in Eq.~\eqref{eq:Estimations} approximates the ``statistical'' variance of the estimator $\widetilde{O}$, and \emph{not} the ``quantum mechanical'' variance of the observable $\langle\hat{O}^2\rangle - \langle\hat{O}\rangle^2$ [indeed Eq.~\eqref{eq:Estimations_Err} calculates the difference between the estimation of the square of the average (first term in the right hand side) and the square of the estimated average (second term)].
A few key observations must be made here. Firstly, in $M$ repeated measurements of $\hat{P}_i$ alone, in agreement with the central limit theorem \cite{shlosberg2023adaptive,fischer2011history} we find in Sec.~\ref{ssec:EstSinglePauli} that $(\Delta\widetilde{O})^2 \overset{M\mapsto \infty}{=} ( \langle \hat{P}_i^2\rangle - \langle \hat{P}_i\rangle^2 ) / M$. 
Furthermore, when two PS $\hat{P}_i$, $\hat{P}_j$ with $i \neq j$, have never been measured together $\widetilde{\theta_i\theta_j} - \widetilde{\theta}_i\widetilde{\theta}_j$ in Eq.~\eqref{eq:Estimations} is zero, and only the scaled variances contribute to the error, as should be the case. 
In this scenario, the posterior that we use in Sec.~\ref{ssec:EstMultPauli} to calculate $\widetilde{\theta_i \theta_j}$ is the product of the posteriors used for $\widetilde{\theta}_i$ and $\widetilde{\theta}_j$ individually, implying $\widetilde{\theta_i \theta_j}=\widetilde{\theta}_i \widetilde{\theta}_j$.

In the following, we explain how $\widetilde{O}$ and $(\Delta\widetilde{O})^2$ are practically determined from the experimental measurement outcomes of both the PS $\hat{P}_i$ and their double $\hat{P}_i \otimes \hat{P}_i$. To do so, we build Bayesian probability distributions \cite{chapman1995bayesian,kruschke2014doing,shlosberg2023adaptive,rick_tbp} to estimate $\widetilde{\theta}_i$, $\widetilde{\theta^2_i}$  and $\widetilde{\theta_i \theta_j}$ and therefore the quantities in Eq.~\eqref{eq:Estimations}. As can be seen in Eqs.~\eqref{eq:Bayes_Estimation} and \eqref{eq:Bayes_Estimation2}, our approach requires distinguishing the cases in which $\widetilde{P}_i$ and $\widetilde{P}_j$ are measured together, alone, or in their double version. 
The procedure for calculating $\widetilde{\theta}_i$ and $\widetilde{\theta_i^2}$, conceptually simpler, is given in Sec.~\ref{ssec:EstSinglePauli}. Sec.~\ref{ssec:EstMultPauli} explains how to find $\widetilde{\theta_i \theta_j}$ when $i \neq j$ and $\hat{P}_i$ commutes with $\hat{P}_j$.

\subsection{Settings and estimation of $\widetilde{\theta}_i$ and $\widetilde{\theta_i^2}$}
\label{ssec:EstSinglePauli}

As depicted in Fig.~\ref{fig:schematics}(b), for an input state $\ket{\varphi}$ and an observable $\hat{O}$ to be measured, the double scheme requires preparing $\ket{\psi} = \ket{\varphi}\otimes\ket{\varphi}$ and simultaneously probing all $\hat{P}_i \otimes \hat{P}_i$ (where $\hat{P}_i$ is in $\hat{O}$).
Calling $\theta_i$ and $\phi_i$ the probabilities of obtaining $+1$ when measuring $\hat{P}_i$ and $\hat{P}_i \otimes \hat{P}_i$, respectively, their relationship is
\begin{equation}\label{eq:phi_i}
\small
    \begin{rcases}
        \bra{\varphi}\hat{P}_i\ket{\varphi} 
        \hspace{-0.8em}&= 
        2\theta_i - 1
        \\
        \underbrace{
        \bra{\psi} \hat{P}_i \otimes \hat{P}_i \ket{\psi}
        }_{
        \bra{\varphi}\hat{P}_i\ket{\varphi}^2
        } 
        \hspace{-0.8em}&= 
        2\phi_i-1
    \end{rcases}
    \Rightarrow
    \phi_i = \theta_i^2+(1-\theta_i)^2.
\end{equation}
This equation confirms the intuitive fact that obtaining $+1$ [$-1$] in the measurement of $\hat{P}_i \otimes \hat{P}_i$ is equivalent to obtaining $(+1,+1)$ or $(-1,-1)$ [$(+1,-1)$ or $(-1,+1)$] in two independent measurements of $\hat{P}_i$.

As can be understood from Eq.~\eqref{eq:phi_i} and Fig.~\ref{fig:schematics}(a), measuring the double PS $\hat{P}_i \otimes \hat{P}_i$ enhances the estimation's accuracy of the absolute values of $\widetilde{P}_i$ in Eq.~\eqref{eq:Estimations}, yet provides no information about their signs -- see top row in Fig.~\ref{fig:schematics}(a): the distribution is symmetric around $1/2$. Therefore, our efficient measurement protocol must distribute the shots between the double scheme $\hat{P}_i \otimes \hat{P}_i$ (that targets \emph{all} $\lvert \widetilde{P}_i \rvert$) and the single scheme $\hat{P}_i$ (where every shot helps determine the signs and magnitudes of \emph{some} $\widetilde{P}_i$). Next, we derive the probability distribution function $\mathcal{P}_1$ to estimate $\widetilde{\theta}_i$ and $\widetilde{\theta_i^2}$ from the measurement outcomes. Alongside $\mathcal{P}_2$ (see Sec.~\ref{ssec:EstMultPauli}), $\mathcal{P}_1$ will be used by our algorithm [see Sec.~\ref{sec:Algo} and Fig.~\ref{fig:schematics}(b)] for the error $(\Delta \widetilde{O})^2$ in Eq.~\eqref{eq:Estimations}, and hence the measurement allocation.

Let us denote $s_{i\pm}$ and $d_{i\pm}$ the number of $\pm 1$ outcomes obtained when probing $\hat{P}_i$ and $\hat{P}_i \otimes \hat{P}_i$, respectively. Following the standard Bayesian approach \cite{kruschke2014doing,chapman1995bayesian}, the expected value $\widetilde{f}$ of any function $f$ of probabilities $\theta_i$ can be determined from the posterior probability $\mathcal{P}_1(\theta_i|s_{i\pm},d_{i\pm})$ by 
\begin{subequations}\label{eq:Bayes_Estimation}
\small
    \begin{align}
        \hspace{-0.65em} \widetilde{f}(\theta_i)
        &=
        \int_0^1 f(\theta_i)\mathcal{P}_1(\theta_i|s_{i\pm},d_{i\pm}) d\theta_i
        , \label{eq:Bayes_Estimation_exp}
        \\ 
        \hspace{-0.65em} \mathcal{P}_1(\theta_i|s_{i\pm},d_{i\pm})
        &\propto
        \theta_{i}^{s_{i +}}(1-\theta_{i})^{s_{i -}}
        \phi_{i}^{d_{i +}}(1-\phi_{i})^{d_{i -}}
        .\label{eq:Bayes_Estimation_post}
    \end{align}
\end{subequations}
Here, we assumed a constant prior for $\mathcal{P}_1$, and that the posterior is normalized: $\int_0^1 \mathcal{P}_1(\theta_i|s_{i\pm},d_{i\pm}) d\theta_i =1$. Furthermore, $\phi_i$ is related to $\theta_i$ through Eq.~\eqref{eq:phi_i}. 

Using Eqs.~\eqref{eq:Bayes_Estimation}, one can substitute $f \mapsto \theta_i $ [$f \mapsto \theta_i^2$] to calculate $(2\widetilde{\theta}_i-1)$ [$(\widetilde{\theta_i^2}-\widetilde{\theta}_i^2)$], as required by Eq.~\eqref{eq:Estimations} to determine $\widetilde{O}$ [$(\Delta\widetilde{O})^2$].

Looking at the features of $\mathcal{P}_1$ in Eq.~\eqref{eq:Bayes_Estimation_post} helps to understand the advantages of our double scheme and the logic behind the allocation algorithm in Sec.~\ref{sec:Algo} and Fig.~\ref{fig:schematics}(b). As can be seen in Fig.~\ref{fig:schematics}(a), when $\hat{P}_i$ has never been measured alone ($s_{i+} = s_{i-} = 0$), its associated $\mathcal{P}_1$ often\footnote{
The distance $\lvert 2\theta_i - 1 \rvert$ between the two peaks must be larger than their widths $\propto (d_{i+} + d_{i-})^{-x}$, $x\in \left[ \frac{1}{4} , 1 \right]$ (depending on $\langle \hat{P}_i \rangle$).
} 
exhibits two symmetric peaks around $\theta_i = 1/2$. This reflects the sign uncertainty that characterizes the double scheme [see Eq.~\eqref{eq:phi_i}]. As long as $d_{i+} + d_{i-}$ is large enough, very few direct measurements of $\hat{P}_i$ suffice to lift this sign uncertainty, collapse one of the peaks, and obtain an accurate estimate for $\widetilde{P}_i$ [top to bottom, Fig.~\ref{fig:schematics}(a)]. As discussed in Sec.~\ref{sec:Results}, our measurement algorithm employs the double scheme to gain information about the magnitudes $\lvert \widetilde{P}_i \rvert$ of all PS, and allocates few extra measurements to discern the signs of $\widetilde{P}_i$ and further enhance their estimation accuracies. 

\subsection{Estimation of $\widetilde{\theta_i \theta_j}$}
\label{ssec:EstMultPauli}

The Bayesian estimation of $\widetilde{\theta_i \theta_j}$ is conceptually similar to the process outlined in the previous Sec.~\ref{ssec:EstSinglePauli}.
Here, we focus on the main differences, arising from the fact that we have data from multiple random variables that must be accounted for simultaneously. Namely, we need to determine a posterior $\mathcal{P}_2$ that describes the cases of two commuting PS $\hat{P}_i$ and $\hat{P}_j$ that can be measured {\color{red}(a)} simultaneously in the double scheme, {\color{blue}(b)} simultaneously in the single scheme, or {\color{green}(c)} individually\footnote{\label{foot:2}
Indeed, in the single scheme we probe simultaneously groups of commuting PS. Say that $\hat{P_i}$ and $\hat{P_j}$ commute. Then, in some groups, they belong together and when measured their outcomes are correlated. However, it is possible (and likely) that there are groups in which only one of them belongs. Suppose $\hat{P}_k$ commutes with $\hat{P}_i$ but not $\hat{P}_j$. Then, there is a group which includes $\hat{P}_k$ and $\hat{P}_i$, but not $\hat{P}_j$. When that group is probed, we obtain uncorrelated (from $\hat{P}_j$) outcomes for $\hat{P}_i$, that must be distinguished from the previous, correlated ones. Hence, the existence of scenarios {\color{blue}(b)} and {\color{green}(c)}.}.

Since the outcomes of two PS can be correlated only if they are measured together, whenever $[\hat{P}_i , \hat{P}_j] \neq 0$ we have $\widetilde{\theta_i \theta_j} = \widetilde{\theta}_i \widetilde{\theta}_j$ and therefore $(\widetilde{\theta_i\theta_j} - \widetilde{\theta}_i \widetilde{\theta}_j) = 0$ in Eq.~\eqref{eq:Estimations}\footnote{
This follows from the fact that when $s_{i\pm j\pm} = 0$ for all combinations $\pm \pm$, $\mathcal{P}_2(\vec{\theta}|s_{i\pm},s_{j\pm},s_{i\pm j\pm},d_{i\pm j\pm}) = \mathcal{P}_1(\theta_i|s_{i\pm},d_{i\pm}) \mathcal{P}_1(\theta_j|s_{j\pm},d_{j\pm})$, see Eqs.~\eqref{eq:Bayes_Estimation_post} and \eqref{eq:Bayes_Estimation2_post}.
}.
As such, we can compute $\widetilde{\theta_i \theta_j}$ working in the simultaneous eigenbasis of $\hat{P}_i$ and $\hat{P}_j$. We denote the probability of the outcome $(\pm1,\pm1)$ when probing $\hat{P}_i$ and $\hat{P}_j$ [$\hat{P}_i \otimes \hat{P}_i$ and $\hat{P}_j \otimes \hat{P}_j$] with $\theta_{i\pm j\pm}$ [$\phi_{i\pm j\pm}$]. By enumerating all possible measurement outcomes, we find relations between $\theta_{i\pm j\pm}$ and $\phi_{i\pm j\pm}$ similar to Eq.~\eqref{eq:phi_i}:
\begin{subequations}\label{eq:phi_ij}
\begin{align}
    \phi_{i+ j+}&=\theta_{i+ j+}^2+\theta_{i+ j-}^2+\theta_{i- j+}^2+\theta_{i- j-}^2,\\
    \phi_{i+ j-}&=2\theta_{i+ j+}\theta_{i+ j-}+2\theta_{i- j+}\theta_{i- j-},\\
    \phi_{i- j+}&=2\theta_{i+ j+}\theta_{i- j+}+2\theta_{i+ j-}\theta_{i- j-},\\
    \phi_{i- j-}&=2\theta_{i+ j+}\theta_{i- j-}+2\theta_{i+ j-}\theta_{i- j+}.
\end{align}
\end{subequations}
From Eqs.~\eqref{eq:phi_ij}, it follows that normalization of $\theta_{i\pm j\pm}$ leads to normalization of $\phi_{i\pm j\pm}$. 

As mentioned in the beginning of this section and in Fig.~\ref{fig:schematics}(b), we distinguish the cases in which $\hat{P}_i$ and $\hat{P}_j$ are measured with the double scheme {\color{red}(a)} and without, and for the latter scenario when they are simultaneously {\color{blue}(b)} or independently {\color{green}(c)} probed (see footnote \ref{foot:2}). We denote $d_{i\pm,j\pm}$ [$s_{i\pm,j\pm}$] \{$s_{i\pm}$ and $s_{j\pm}$\} the number of $(\pm1,\pm1)$ [$(\pm1,\pm1)$] \{$\pm1$ for $\hat{P}_i$ and $\pm1$ for $\hat{P}_j$\} outcomes in these three cases, {\color{red}(a)}, [{\color{blue}(b)}], and \{{\color{green}(c)}\}, respectively\footnote{
Notice that $s_{i\pm}$ ($s_{j\pm}$) in Eqs.~\eqref{eq:Bayes_Estimation2}, in agreement with case {\color{green}(c)}, only includes outcomes where $\hat{P}_i$ ($\hat{P}_j$) has been measured independently from $\hat{P}_j$ ($\hat{P}_i$). This is in contrast to Eqs.~\eqref{eq:Bayes_Estimation}, where $s_{i\pm}$ includes all outcomes obtained for $\hat{P}_i$.
}.
Using $\vec{\theta} = \{ \theta_{i+ j+} , \theta_{i- j+} , \theta_{i+ j-} \}$, Eqs.~\eqref{eq:Bayes_Estimation} can be generalized to yield the estimate $\widetilde{g}$ of an arbitrary function $g$ of $\vec{\theta}$ by
\begin{subequations}\label{eq:Bayes_Estimation2}
\small
    \begin{align}
        &
        \widetilde{g}(\vec{\theta})
        =
        \int g(\vec{\theta})\mathcal{P}_2(\vec{\theta}|s_{i\pm},s_{j\pm},s_{i\pm j\pm},d_{i\pm j\pm}) d\vec{\theta}
        ,
        \label{eq:Bayes_Estimation2_exp}
        \\
        &
        \begin{aligned}
        \mathcal{P}_2
        (
        &
        \vec{\theta}|s_{i\pm},s_{j\pm},s_{i\pm j\pm},d_{i\pm j\pm})
        \propto
        &
        \\
        &
        \phi_{i+j+}^{d_{i+j+}} \phi_{i+j-}^{d_{i+j-}} \phi_{i-j+}^{d_{i-j+}} \phi_{i-j-}^{d_{i-j-}}
        &
        \leftarrow
        \text{{\color{red}(a)}}
        \\
        &
        \theta_{i+j+}^{s_{i+j+}} \theta_{i+j-}^{s_{i+j-}} \theta_{i-j+}^{s_{i-j+}} \theta_{i-j-}^{s_{i-j-}}
        &
        \leftarrow
        \text{{\color{blue}(b)}}
        \\
        &
        \theta_{i}^{s_{i +}}(1-\theta_{i})^{s_{i -}}
        \theta_{j}^{s_{j +}}(1-\theta_{j})^{s_{j -}}
        &
        \leftarrow
        \text{{\color{green}(c)}}
        \end{aligned}
        \label{eq:Bayes_Estimation2_post}
    \end{align}
\end{subequations}
where $\phi_{i\pm j\pm}$ is given by Eq.~\eqref{eq:phi_ij}, $\theta_i = \theta_{i+j+} + \theta_{i+j-}$ and $\theta_j = \theta_{i+j+} + \theta_{i-j+}$.

Similarly to Eq.~\eqref{eq:Bayes_Estimation_post}, Eq.~\eqref{eq:Bayes_Estimation2_post} must be normalized: $\int \mathcal{P}_2(\vec{\theta}|s_{i\pm},s_{i\pm j\pm},d_{i\pm j\pm}) d\vec{\theta} =1$. Here and in Eq.~\eqref{eq:Bayes_Estimation2_exp}, the integral is to be taken over three independent probabilities within $\vec{\theta}$ such that their sum never exceeds unity. We note that the posterior $\mathcal{P}_2$ in Eq.~\eqref{eq:Bayes_Estimation2_exp} is the product of the individual posteriors {\color{red}(a)}, {\color{blue}(b)} and {\color{green}(c)}, as these experiments are carried out independently.

By plugging 
\begin{equation}
    \begin{split}
        g 
        \mapsto 
        &
        \, \theta_i\theta_j = (\theta_{i+ j+} + \theta_{i+ j-})(\theta_{i+ j+} + \theta_{i- j+})
    \end{split}
\end{equation}
into Eqs.~\eqref{eq:Bayes_Estimation2}, together with the results from Sec.~\ref{ssec:EstSinglePauli}, we find $\widetilde{\theta_i \theta_j}$ and can determine $(\Delta \widetilde{O})^2$ as described in Eq.~\eqref{eq:Estimations_Err}. This quantity will be central to our algorithm in Sec.~\ref{sec:Algo}. 

\section{Algorithm}
\label{sec:Algo}
In this section, we explain the algorithm that allocates the $M$ shots available to the measurement. Each time, it will choose between simultaneously probing all double PS $\hat{P}_i \otimes \hat{P}_i$ or a group of commuting PS $\hat{P}_i$, see Fig.~\ref{fig:schematics}(b). Since the double scheme requires twice the number of qubits to measure $\ket{\psi} = \ket{\varphi} \otimes \ket{\varphi}$ compared to $\ket{\varphi}$ ($\ket{\varphi}$ being the state to be measured), it costs twice as many measurement shots, and at least twice the amount of resources. To assess the advantage yielded by the double scheme, we define the effective measurement shots
\begin{equation}
    M_{\rm eff}
    =
    M_{\rm single} + 2\alpha M_{\rm double}
    , \label{eq:M_eff}
\end{equation}
where $M_{\rm double}$ and  $M_{\rm single}$ are shots assigned and not assigned to the double scheme, respectively. This definition of $M_{\rm eff}$ is equivalent to saying that measuring all $\hat{P}_i \otimes \hat{P}_i$ requires $2\alpha$-times the physical resources. Thus, in order to be advantageous, probing all $\hat{P}_i \otimes \hat{P}_i$ \emph{once} must lower the estimation error $(\Delta\widetilde{O})^2$ in Eq.~\eqref{eq:Estimations} more than probing $2\alpha$ groups of commuting PS $\hat{P}_i$ (or the same group $2\alpha$ times). The factor $\alpha \geq 1$ takes into account the additional resources required to simultaneously diagonalize and measure all $\hat{P}_i \otimes \hat{P}_i$ \cite{shlosberg2023adaptive, Jena2019}.
As we will see in Sec.~\ref{sec:Results}, despite this extra weight, the double scheme is often advantageous provided the observable $\hat{O}$ is not trivial (in the sense that the $p$ PS within do not commute with each other).

The pipeline of the algorithm is shown in Fig.~\ref{fig:schematics}(b). Before each measurement, we determine the \emph{expected} improvement to the estimated error $(\Delta\widetilde{O})^2$ depending on whether the shot is assigned to the double scheme or one of the groups containing commuting PS $\hat{P}_i$. This is done by allocating \emph{virtual} measurements that yield outcomes in agreement with the knowledge already acquired. 
In the toy example presented in Fig.~\ref{fig:schematics}(b), our algorithm considers the blue, green, pink, and black (for the double) groups of PS and assesses how their contribution to the error $(\Delta \widetilde{O})^2$ changes when this virtual measurement is assigned to them.

In practice, the virtual measurement is achieved by adding to $s_{i\pm}$ or  $d_{i\pm}$ within $\mathcal{P}_1$ in Eq.~\eqref{eq:Bayes_Estimation_post} [$s_{i\pm}$, $s_{j\pm}$, $s_{i\pm j\pm}$ or $d_{i\pm j\pm}$ within $\mathcal{P}_2$ in Eq.~\eqref{eq:Bayes_Estimation2_post}] the expected outcome (based on the current knowledge). This is equivalent to a change in the prior, effectively positing that, through a virtual measurement, the outcomes are expected to change according to our prior observations. An heuristic that does this prescribes the following substitutions:
%

\begin{subequations}
\label{eq:virtual_msmnt}
    \begin{align}
    \hspace{-1em}&
        \begin{cases}
            d_{i+} 
            \mapsto 
            d_{i+} + \frac{\widetilde{\phi}_i}{2\alpha}
            \\
            d_{i-} 
            \mapsto 
            d_{i-} + \frac{1-\widetilde{\phi}_i}{2\alpha}
        \end{cases}
        &
        \hspace{-0.5em}
        \begin{array}{l}
             \text{with } \widetilde{\phi}_i  \\
             \text{from Eq.~\eqref{eq:Bayes_Estimation},} 
        \end{array}
        \label{eq:virtual_msmnt_double_1}
        \\
        \hspace{-1em}&
        d_{i\pm j\pm} 
        \mapsto 
        d_{i\pm j\pm} + \frac{\widetilde{\phi}_{i\pm j\pm}}{2\alpha}
        &
        \hspace{-0.5em}
        \begin{array}{l}
             \text{with } \widetilde{\phi}_{i\pm j\pm}  \\
             \text{from Eq.~\eqref{eq:Bayes_Estimation2},} 
        \end{array} \label{eq:virtual_msmnt_double_2}
        \\
        \hspace{-1em}&
        \begin{cases}
            s_{i+} 
            \mapsto 
            s_{i+} + \widetilde{\theta}_i
            \\
            s_{i-} 
            \mapsto 
            s_{i-} + (1-\widetilde{\theta}_i)
        \end{cases}
        &
        \hspace{-0.5em}
        \begin{array}{l}
             \text{with } \widetilde{\theta}_i  \\
             \text{from Eq.~\eqref{eq:Bayes_Estimation},} 
        \end{array} \label{eq:virtual_msmnt_single_1}
        \\
        \hspace{-1em}&
        s_{i\pm j\pm} 
        \mapsto 
        s_{i\pm j\pm} + \widetilde{\theta}_{i\pm j\pm}
        &
        \hspace{-0.5em}
        \begin{array}{l}
             \text{with } \widetilde{\theta}_{i\pm j\pm}  \\
             \text{from Eq.~\eqref{eq:Bayes_Estimation2},} 
        \end{array}\label{eq:virtual_msmnt_single_2}
    \end{align}
\end{subequations}
where $\widetilde{\theta}_i$ and $\widetilde{\phi}_i$ [$\widetilde{\theta}_{i\pm j\pm}$ and $\widetilde{\phi}_{i\pm j\pm}$] are obtained from $f(\theta_i) \mapsto \theta_i$ and $f(\theta_i) \mapsto \phi_i$ in Eq.~\eqref{eq:Bayes_Estimation_exp} [$g(\vec{\theta}) \mapsto \theta_{i\pm j\pm}$ and $g(\vec{\theta}) \mapsto \phi_{i\pm j\pm}$ in Eq.~\eqref{eq:Bayes_Estimation2_exp}], respectively, and making use of Eq.~\eqref{eq:phi_i} [Eqs.~\eqref{eq:phi_ij}].
Furthermore, the factors $2\alpha$ dividing $\widetilde{\phi}_i$, $(1-\widetilde{\phi}_i)$, and $\widetilde{\phi}_{i\pm j\pm}$ in Eqs.~\eqref{eq:virtual_msmnt_double_1} and \eqref{eq:virtual_msmnt_double_2} take into account the extra resources required by the double scheme (see Eq.~\eqref{eq:M_eff}).

Once the \emph{expected} improvements for all possible choices (all $\hat{P}_i \otimes \hat{P}_i$ for the double scheme or any group of commuting PS $\hat{P}_i$) available to the algorithm are determined, we assign the measurement shot. The choice that leads to the greatest improvement, that is, the smallest resulting error $(\Delta\widetilde{O})^2$ in Eq.~\eqref{eq:Estimations}, will be the one taken by our algorithm. For the toy example in Fig.~\ref{fig:schematics}(b), that is the green group. Ideally, this should be done at each of the $M$ steps of the process; in practice, however, it is possible to improve the resource scaling by updating the error prediction only intermittently once we accept minor compromises in terms of obtained errors (see next paragraph). Taking into account the Markov Chain Monte Carlo process (see Refs.~\cite{Newman1999, rick_tbp, Brooks2011} for details) employed to estimate $\widetilde{\theta}_i$ and $\widetilde{\theta_i \theta_j}$ in Eq.~\eqref{eq:Estimations_Err} ($p^2/2$ terms that need to be calculated) in Sec.~\ref{sec:Theory}, the scaling of the entire measurement algorithm is at most $\mathcal{O}(p^2 M^3)$ (one factor of $M$ for updating the errors at every shot and $M^2$ for the Monte Carlo process to reach the required precision). We also point out that assigning virtual measurements does not present a bottleneck of the algorithm. From Eq.~\eqref{eq:virtual_msmnt}, it follows that the terms $\widetilde{\theta}_i$, $\widetilde{\theta_i^2}$, and $\widetilde{\theta_i \theta_j}$ in Eq.~\eqref{eq:Estimations_Err} need to be updated four times at most, reflecting the possible cases {\color{red}(a)}, {\color{blue}(b)}, and {\color{green}(c)} (twice for the latter).

\begin{figure*}[ht!]
    \centering
    \includegraphics[width=0.9\textwidth]{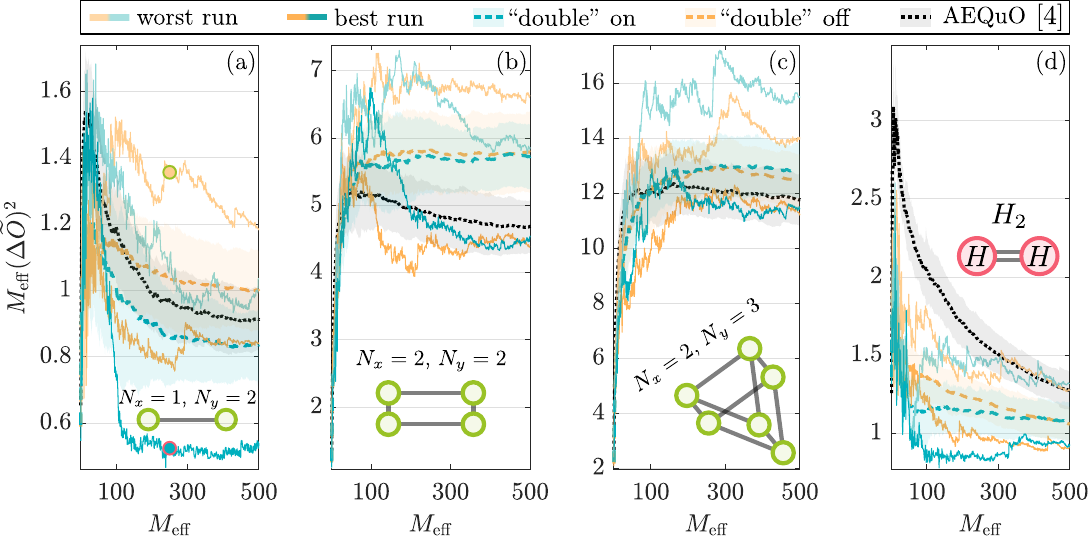}
    \caption{
    Numerical results with Ising (a-c) and chemistry (d) Hamiltonians -- see main text -- and their ground states as input $\ket{\varphi}$. Dashed (our method) and dotted (AEQuO \cite{shlosberg2023adaptive}) lines are the average from $25$ repeated measurement processes, and their shadows correspond to their root mean squares. Blue (Yellow) is used when the double scheme is (not) employed. Dark and light lines are the best and worst runs at large $M_{\rm eff}$, respectively, within the $25$ repetitions. The histograms in Figs.~\ref{fig:histograms}(c) refer to the two dots at $M_{\rm eff} = 250$.
    }
    \label{fig:numerics}
\end{figure*}

The time scaling of the algorithm can be drastically improved since the independent contributions $\widetilde{\theta_i\theta_j} - \widetilde{\theta}_i\widetilde{\theta}_j$ in Eq.~\eqref{eq:Estimations} can be calculated in parallel and one can use multiple Monte Carlo chains for each estimation \cite{Gelman1992,Brooks01121998}. 
Furthermore, instead of assigning shots one-by-one, another viable option is to allocate multiple shots at each step of the algorithm, such that the error contributions in Eq.~\eqref{eq:Estimations_Err} do not have to be computed at each shot. This can be done based on the expected error improvements either deterministically (allocating shots as to mimic the relative size of the errors for each group) or probabilistically (normalizing the calculated errors and using them as a probability distribution). 
Hence, a realistic time scaling sits in between $\mathcal{O}(M)$ and $\mathcal{O}(p^2 M^3)$, depending on the available computational resources and the chosen allocation strategy.

\section{Numerical results}
\label{sec:Results}
In this section, we benchmark our scheme by turning on and off the possibility of allocating measurement to the double scheme in Fig.~\ref{fig:schematics}(b). We therefore unambiguously determine whether it is advantageous to sacrifice some measurement shots [see Sec.~\ref{sec:Algo} and the definition of $M_{\rm eff}$, Eq.~\eqref{eq:M_eff}] to probe all PS $\hat{P}_i \otimes \hat{P}_i$ during the measurement process. Furthermore, we compare the estimation errors $(\Delta \widetilde{O})^2$ obtained through these two methods with the errors yielded by AEQuO \cite{shlosberg2023adaptive} and by a naive strategy. AEQuO is, to our knowledge, the state of the art in measurement protocols and until this work the only approach that provides single run estimates for both the average $\widetilde{{O}}$ and the error $(\Delta \widetilde{O})^2$ of the input observable $\hat{O}$, see Eq.~\eqref{eq:Estimations}. With ``single run'' we mean that the assessment of the error that affects $\widetilde{O}$ does not require repeating the entire measurement process multiple times. Instead, it is produced together with $\widetilde{O}$ in a single run. On the other hand, the naive strategy consists in allocating a fixed percentage of the shots to the double scheme, and randomly assigning the others to groups of commuting PS, weighting each choice with the sum of their coefficients squared -- similar to Refs.~\cite{Huang2021, Crawford2021}.

For our numerical results, we consider two physical systems. The first is the Hydrogen molecule ${\rm H_2}$ in its $6$-$31$G, JW encoding \cite{cao2019quantum,mcclean2020openfermion}, with $q = 8$ qubits and $p = 185$ PS. 
The second is a quantum Ising model of nuclear spins on rectangular lattices with periodic boundary conditions, $N_x$ and $N_y$ vertices along the $x$ and $y$ directions, with higher-order spin-spin tensor interactions \cite{Blanchard2015measurement} included perturbatively. The three instances considered here are shown in the insets in Figs.~\ref{fig:numerics}(a-c) and have $(N_x, N_y)$ equal to $(1,2)$, $(2,2)$, and $(2,3)$, resulting in Hamiltonians $\hat{\mathcal{H}}_{\rm I}$ with $q=2,4,6$ qubits and $p=15,48,99$ PS, respectively. The $\hat{\mathcal{H}}_{\rm I}$ used in this work is
\begin{equation}
\label{eq:ising_ham}
    \begin{aligned}
        &\hat{\mathcal{H}}_{\rm I} 
        = 
        \sum_{i} 
        B_{i}^{\rm z} \hat{Z}_i
        +
        \sum_{\langle i,k\rangle} 
        J_{ik} \hat{Z}_i \hat{Z}_k
        &
        \leftarrow \text{Ising}
        \\
        &+
        \sum_{i} 
        \boldsymbol{b}_i 
        \cdot 
        \boldsymbol{\hat{\sigma}}_i
        +
        \sum_{\langle i,k\rangle} 
        \underline{j_{ik}}
        \cdot
        \boldsymbol{\hat{\sigma}}_i 
        \otimes 
        \boldsymbol{\hat{\sigma}}_k
        ,
        & 
        \leftarrow \text{tensor}
    \end{aligned}
\end{equation}
where $\langle i,k\rangle$ indicates connected vertices $i,k$ in the lattice, $B_{i}^{\rm z}$, $J_{ik}$, $\boldsymbol{b}_i$ and $\underline{j_{ik}}$ are coefficients characterizing the system, and we explicitly divided single- and two-body terms. Bold indicates vectors (e.g., $\boldsymbol{\hat{\sigma}}_i = \lbrace \hat{X}_i , \hat{Y}_i, \hat{Z}_i \rbrace$) and $\underline{j_{ik}}$ is a symmetric matrix for all $i,k$. Numerical values used for the results reported in Fig.~\ref{fig:numerics} are listed in App.~\ref{app:numerics}.

To assess the advantage yielded by the double scheme, we use the effective measurement shots $M_{\rm eff}$ in Eq.~\eqref{eq:M_eff}.
As discussed above, to utilize the double scheme one requires at least twice the resources compared to standard methods. This is also taken into account when we estimate the expected improvement to $(\Delta\widetilde{O})^2$ -- see factors $1/(2\alpha)$ in Eqs.~\eqref{eq:virtual_msmnt} and the related discussion.
In the following, we will assume the ideal case $\alpha = 1$.
We remark, however, that our scheme is implementation ready: it also provides the measurement Clifford circuits for both the single and the double measurements. From these, it is possible to precisely assess the extra resources required to implement the double scheme, and therefore the factor $\alpha$. In practical scenarios, we expect $\alpha \approx 1$, due to the fact that measurement circuits are often shallower or comparable in depth to the algorithmic circuits preceding them.

The numerical results in Fig.~\ref{fig:numerics} suggest that the advantage of the double scheme is highly dependent on the properties of the observable $\hat{O}$ and input state $\ket{\varphi}$. Looking at panels (a-c) -- associated to the Ising Hamiltonians in Eq.~\eqref{eq:ising_ham} -- it is possible to notice that the (rescaled) error $M_{\rm eff} (\Delta\widetilde{O})^2$ obtained with the double scheme (light blue dashed line) is significantly better than the one without (yellow dashed line) only for the case $(N_x = 1,N_y=2)$. This results from the PS $\hat{P}_i$ in the Hamiltonian being less likely to commute between each other compared to the instances $(N_x = 2,N_y=2)$ and $(N_x = 2,N_y=3)$\footnote{
Since the Ising Hamiltonian in Eq.~\eqref{eq:ising_ham} is characterized by single- and two-body interactions, the larger the lattice $(N_x,N_y)$ the more identities within each PS $\hat{P}_i$.
}.

\begin{figure*}[ht!]
    \centering
    \includegraphics[width=\textwidth]{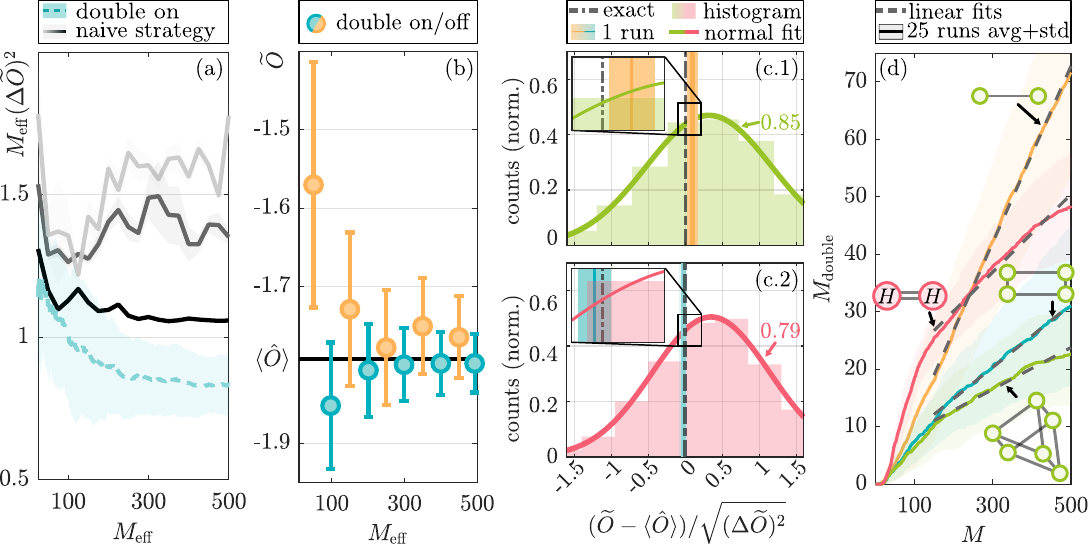}
    \caption{
    (a) Comparison between our protocol with the double (blue) and the naive strategy (grey, see main text); same settings as in Fig.~\ref{fig:numerics}(a). The percentages of shots allocated to measurements without the double scheme is $75 \%$ (light grey), $50 \%$ (grey), and $25 \%$ (dark grey).
    Panels (b) and (c) demonstrate the statistical validity of the estimators $\widetilde{O}$ and $(\Delta \widetilde{O})^2$ in Sec.~\ref{sec:Theory}. (b) reports intermediate values of $\widetilde{O}$ within a single run, with error bars $\sqrt{(\Delta \widetilde{O})^2}$. Blue (Yellow) does (not) employ the double scheme; exact value $\langle \hat{O} \rangle$ is the horizontal black line.
    In (c), we simulate the measurement process $5000$ times with the allocations from the points at $M_{\rm eff} = 250$ highlighted in Fig.~\ref{fig:numerics}(a), and collect the values $
    (\widetilde{O}-\langle \hat{O} \rangle)
    /
    \sqrt{(\Delta\widetilde{O})^2}
    $ in the histograms. Full lines are Gaussian fits; their root mean squares reported. Vertical full lines and their shadows (visible in the insets) are the averages $\widetilde{O} - \langle \hat{O} \rangle$ and errors $\sqrt{(\Delta\widetilde{O})^2}$ (respectively) extracted from the points highlighted in Fig.~\ref{fig:numerics}(a). 
    For (a-c), we used the Ising Hamiltonian with $N_x = 2$ and $N_y =1$ and the ground state as input $\ket{\varphi}$.
    In (d), we report $M_{\rm double}$ [see discussion below Eq.~\eqref{eq:M_eff}] as a function of $M$ for all instances investigated in Fig.~\ref{fig:numerics}. Dashed lines are linear fits, with slopes $0.154$, $0.055$, $0.036$ [Ising Hamiltonians in Figs.~\ref{fig:numerics}(a-c)] and $0.068$ [chemistry molecule in Fig.~\ref{fig:numerics}(d)].
    }
    \label{fig:histograms}
\end{figure*}

This conclusion is reinforced by the results in Fig.~\ref{fig:numerics}(d), where the observable is the Hydrogen molecule ${\rm H_2}$. 
Molecular Hamiltonians are characterized by many-body interactions such that several of their PS $\hat{P}_i$ have few identities. This means that the double scheme is more advantageous and, for small $M_{\rm eff}$ (see below why), yields better errors $(\Delta\widetilde{O})^2$. Generally, the more likely it is for PS $\hat{P_i}$ to be non-commuting, and the more PS $p$ there are in the decomposition of $\hat{O}$ in Eq.~\eqref{eq:Estimations}, the more advantageous the double scheme is. This is further supported by the number of times $M_{\rm double}$ our algorithm chooses to employ the double scheme, see Fig.~\ref{fig:histograms}(d) and discussion below.

An insightful comparison to be made is between AEQuO \cite{shlosberg2023adaptive} (black dotted line) and our schemes, with and without the double feature activated. AEQuO is, to our knowledge, the state of the art when it comes to estimating observables on quantum computers while providing statistical errors $(\Delta\widetilde{O})^2$ within a single run. It also allocates measurements based on the instantaneous knowledge it has on the estimated variance $(\Delta \widetilde{O})^2$, although that knowledge is less precise than ours~\cite{rick_tbp} and it cannot access the double scheme.
As can be seen in Fig.~\ref{fig:numerics}(a), sometimes AEQuO yields better errors $(\Delta\widetilde{O})^2$, despite a lower estimation accuracy of the covariances and equivalent terms $\widetilde{\theta_i \theta_j} - \widetilde{\theta}_i\widetilde{\theta}_j$ in Eq.~\eqref{eq:Estimations}, see Ref.~\cite{rick_tbp}. We believe that this results from different strategies in the shot allocation due to the instantaneous knowledge available at each step. A similar behaviour was observed in Ref.~\cite{shlosberg2023adaptive} in the comparison of machine-learning and bucket filling approaches. We remark, however, that in the scenarios where the double scheme is advantageous (many non-commuting PS $\hat{P}_i$, Figs.~\ref{fig:numerics}(a,d), see above), the double scheme performs considerably better than AEQuO.

Beyond the comparison with AEQuO, it is interesting to understand how well our adaptive allocation scheme works. In Fig.~\ref{fig:histograms}(a) we benchmark our double scheme, which allocates shots one by one to the most promising group of commuting PS or all double PS (see Fig.~\ref{fig:schematics}(b) and Sec.~\ref{sec:Algo}), against a naive strategy that assigns fixed percentages $p_{\rm double}$ and $1-p_{\rm double}$ of the measurement budget to the double scheme and to groups of commuting PS, respectively. The three grey lines in the figure are obtained for $p_{\rm double}$ equal to $0.25$ (darkest), $0.5$ and $0.75$ (lightest). This means that, taking $M_{\rm eff} = 100$ and $p_{\rm double} = 0.5$ as an example, the naive strategy assigns $25$ shots to the double scheme and $50$ to groups of commuting observables (see Eq.~\eqref{eq:M_eff} with $\alpha = 1$). For the latter, it chooses randomly and non-adaptively (meaning that it does not refine the selection based on previously gathered knowledge, see Sec.~\ref{sec:Algo}), preferring groups whose sums of squares of the PS coefficients $\sum_i c_i^2$ are larger \cite{Huang2021, Crawford2021}\footnote{After allocating all shots and performing the measurements, we check whether any group of commuting PS has a negative error contribution, similarly to \cite{shlosberg2023adaptive}, and in that case discard the result. These negative errors, clearly unphysical, arise from insufficient statistics for the sign estimation, and occur for large values of $p_{\rm double}$ (our adaptive allocation ensures the signs are estimated correctly before the double scheme is employed). We are currently working to further enhancing the Bayesian estimation framework \cite{rick_tbp, shlosberg_tbp} to avoid these cases and ensure better estimation overall.}. 

As can be seen in the figure, our adaptive approach yields significantly better results. Compared to the naive strategies with $p_{\rm double} = 0.25$, $0.50$ and $0.75$, at $M_{\rm eff} = 500$ our approach yields variances $(\Delta \widetilde{O})^2$ that are on average $1.3$, $1.6$, and $2.1$ times smaller, respectively.
Unsurprisingly, the naive strategy that yields better results is the one with $p_{\rm double} = 0.25$, which approximates best the number of shots assigned to the double scheme by our adaptive approach, see Fig.~\ref{fig:histograms}(d). As discussed below, the optimal number of times to employ the double scheme is highly problem dependent. This works in favour of our adaptive approach, where instead of using fixed percentages, it chooses what to measure at each step depending on the available knowledge at that time.
An interesting observation is that the naive strategy with $p_{\rm double} =0.25$ is more stable than the others with higher $p_{\rm double} = 0.5$ and $0.75$. This can be understood from the shaded area (representative of the variations between runs) and the smaller fluctuations of the average. We believe that this is a consequence of the fact that, with fewer shots assigned to groups of PS, the signs of individual PS are less certain for larger $p_{\rm double}$.

An important question is whether the error $(\Delta\widetilde{O})^2$ in Eq.~\eqref{eq:Estimations} is representative of the statistical uncertainty of the estimate $\widetilde{O}$. This is investigated in the scatter plot and the histograms in Figs.~\ref{fig:histograms}(b,c), respectively. The first collects single run estimates $\widetilde{O}$ from our approach with (blue) and without (yellow) the double scheme. Error bars are $\sqrt{(\Delta \widetilde{O})^2}$ and, except for the first point at $M_{\rm eff} = 25$, all intersect the black full line that represents $\langle \hat{O} \rangle$. This suggests that our estimates for both the average and its error are statistically meaningful. The latter, in particular, is not over- or underestimated.

To further confirm this, the histograms in Figs.~\ref{fig:histograms}(c) collect $5000$ values of 
\begin{equation}
    \frac{
    \widetilde{O}-\langle \hat{O}\rangle
    }{
    \sqrt{
    (\Delta\widetilde{O})^2
    }
    }
\end{equation}
obtained in repeated measurement processes with a budget of $M_{\rm eff} = 250$. For each of those, $\widetilde{O}$ and $(\Delta\widetilde{O})^2$ are calculated using Eq.~\eqref{eq:Estimations} and the methodology described in Sec.~\ref{sec:Theory}. As can be seen in the figure, both histograms with (pink) and without (green) the double scheme are centred nearby zero (black dash-dotted vertical line) and have a root mean square (numbers reported in the plots) that is just below unity. This follows from the Bayesian procedure in Sec.~\ref{sec:Theory}. For small measurement budgets $M_{\rm eff}$, the chosen priors ensure that $(\Delta\widetilde{O})^2$ is overestimated to compensate for biases \cite{kruschke2014doing,chapman1995bayesian,rick_tbp} (due to the non-commutativity of the PS $\hat{P}_i$, for $M_{\rm eff} = 250$ some PS are measured approximately ten times). For reference, the vertical, full lines in Figs.~\ref{fig:histograms}(c) are the averages from the corresponding data points on the single runs highlighted in Fig.~\ref{fig:numerics}(a), and their shadows cover $\pm\sqrt{(\Delta\widetilde{O})^2}$. 
This illustrates that with or without the double scheme, our Bayesian framework correctly reproduces the statistical error. In other words, employing $(\Delta \widetilde{O})^2$ to assess the quality of our estimator is statistically equivalent to $\sum_i(\widetilde{O}_i - \langle \hat{O} \rangle)^2 / N_{\rm runs}$, i.e. the variance obtained from $N_{\rm runs}$ repeated evaluations $\widetilde{O}_i$ of $\langle \hat{O} \rangle$, with the differences that (a) we do not need to perform repeated runs and (b) we do not rely on the exact expectation value $\langle \hat{O} \rangle$, which is generally unknown.

In Fig.~\ref{fig:histograms}(d) we show $M_{\rm double}$ as a function of $M$ for all instances considered in Fig.~\ref{fig:numerics}. A few characteristics are common to all lines. First, our algorithm [Sec.~\ref{sec:Algo} and Fig.~\ref{fig:schematics}(b)] starts with an exploratory phase, where the signs of the PS $\hat{P}_i$ are assessed. This corresponds to the flat part of the lines and avoids situations in which the posterior $\mathcal{P}_1$ presents two peaks [see Fig.~\ref{fig:schematics}(a)], resulting in large error contributions. Second, the assignment of shots to the double scheme scales approximately linearly with $M$. This is a consequence of the sign uncertainty that limits the accuracy after repeatedly measuring with the double scheme. The slopes of the grey dashed lines in Fig.~\ref{fig:histograms}(d) mainly depend on the non-commuting terms in the observable. With more, only a small fraction of the PS $\hat{P}_i$ is probed when the double scheme is not used, and many shots must be employed to collect signs' information. This makes the double scheme more convenient, as indicated by the steeper linear fit of the $(N_x=1,N_y=2)$ Ising (yellow line) Hamiltonian. To support this observation, note that the yellow, blue, and green lines (referring to the Ising Hamiltonians) are all flat until around $M \approx 15$. This suggests that (despite larger lattices $(N_x,N_y)$ have more PS $p$) measuring all PS $\hat{P}_i$ can be done with a similar number of measurements in all Ising instances.

Finally, we draw attention to the results obtained for the chemistry molecules in Figs.~\ref{fig:numerics}(d) and \ref{fig:histograms}(d). As can be seen in the latter, the double scheme is chosen more often when the number of shots $M$ is between $20$ and $150$. Indeed, the slope of the pink like is suddenly reduced for larger $M \gtrsim 150$. Coincidentally, the errors $(\Delta \widetilde{O})^2$ obtained with the double scheme are considerably better within this regime, see Fig.~\ref{fig:numerics}(d). This is a consequence of the fact that PS whose averages are approximately zero are less convenient to measure with the double scheme \cite{King2024triply}, as their error contribution $(\Delta \widetilde{O})^2$ scales $\propto 1/\sqrt{M}$ rather than $\propto 1/M$. In fact, due to the square in Eqs.~\eqref{eq:phi_i}, the accuracy enhancement from the double scheme depends on the expectation values of the PS. The larger $\langle \hat{P}_i \rangle $ is in absolute value, the better. 
This can be understood from the probability distribution in the top row of Fig.~\ref{fig:schematics}(a), where the number of single measurements $s_{i+} + s_{i-}$ is $0$. When the peaks are distinct, which happens for sufficiently many measurements and an average of $\langle \hat{P}_i \rangle$ that is large enough, we get two individual peaks, each with variance $\propto 1/M$ as if we measured only one Pauli with the single measurement scheme. However, with the peaks approaching each other, the variance increases until it becomes $\propto 1/\sqrt{M}$ when $\langle \hat{P}_i \rangle = 0$.
In the chemistry example, at around $M \approx 150$ the error contributions from PS with low expectation values start to be important, and therefore the double scheme becomes less effective.

\section{Conclusions}
\label{sec:Conclusions}

We presented an application-ready, multi-copy measurement protocol \cite{Huang2022quantum,chen2021exponential,Chen2024optimal,King2024triply} that we call the double measurement scheme, based on a Bayesian enhanced framework that exploits joint PS measurements to improve the estimation of observables on quantum computers. Our algorithm adaptively allocates future measurements based on instantaneous knowledge, yields estimates for both the average and the error of the desired quantity, and exploits general commutation relations between PS. Our results show that the double scheme is often advantageous, in the sense that it requires fewer shots to reach the desired precision. Notably, when the observable is characterized by many non-commuting Pauli terms, the double scheme performs particularly well and greatly outperforms AEQuO \cite{shlosberg2023adaptive}.

As a first outlook, it would be highly desirable to adapt the code to run on GPU architectures. The resulting acceleration would greatly improve the scaling of the algorithm~\cite{Terenin2019GPUaccelerated}, thus allowing us to tackle larger observables for higher values of shots $M$. These regimes, with more non-commuting PS, are expected to be more advantageous for the double scheme than the small-scale instances considered in this work.
Secondly, we want to extend the double scheme to incorporate sign information more directly. This could be achieved by ancillary systems \cite{Markovich2025quantum}, with the caveat of more complex single shot measurements. It would also be valuable to analyse the performance of this approach in the presence of hardware noise \cite{rick_tbp}, and to integrate it with error-mitigation techniques. Finally, the concept of double measurement could be generalized beyond Pauli observables to other operator families or even continuous variable systems, as well as other measurement schemes that use different posteriors. 

\section*{Author Contributions}
All authors contributed to the manuscript. This work was developed by R.S based on previous work done by A.J., with analytical support from Z.S. and numerics carried out by C.N. under the guidance and direction of L.D. Generative AI was not used in the construction of this manuscript and was at most used sparingly as a resource to look up information otherwise hard to access. 

\section*{Acknowledgments}
We thank Margaritis Voliotis for fruitful probabilistic discussions. LD, CN, and RS acknowledge the EPSRC quantum
career development grant EP/W028301/1 and the
EPSRC Standard Research grant EP/Z534250/1. We would also like to thank the reviewers for their insightful comments and questions that elevated the manuscript considerably.

\bibliographystyle{unsrtnat}
\bibliography{literature.bib}

\clearpage
\appendix

\section{Numerical simulations}
\label{app:numerics}
In this section, we include the randomly generated coefficients $c_i$ and PS $\hat{P}_i$ for the Ising Hamiltonians, Eq.~\eqref{eq:ising_ham} and $(N_x, N_y)$ equal to $(1,2)$ [table~\ref{tab:ising_21}], $(2,2)$ [table~\ref{tab:ising_22}], and $(2,3)$ [table~\ref{tab:ising_23}].

\begin{table}[h!]
\footnotesize
\centering
\begin{tabular}{|c|c|}
\hline
\textbf{Coefficient $c_i$} & \textbf{PS $\hat{P}_{i}$} \\
\hline\hline
-0.005 & $\hat{X} \hat{I}$ \\
\hline
-0.044 & $\hat{Y} \hat{I}$ \\
\hline
1.028 & $\hat{Z} \hat{I}$ \\
\hline
0.198 & $\hat{I} \hat{X}$ \\
\hline
-0.053 & $\hat{I} \hat{Y}$ \\
\hline
1.100 & $\hat{I} \hat{Z}$ \\
\hline
0.121 & $\hat{X} \hat{X}$ \\
\hline
-0.022 & $\hat{X} \hat{Y}$ \\
\hline
-0.104 & $\hat{X} \hat{Z}$ \\
\hline
-0.022 & $\hat{Y} \hat{X}$ \\
\hline
0.034 & $\hat{Y} \hat{Y}$ \\
\hline
-0.143 & $\hat{Y} \hat{Z}$ \\
\hline
-0.104 & $\hat{Z} \hat{X}$ \\
\hline
-0.143 & $\hat{Z} \hat{Y}$ \\
\hline
0.416 & $\hat{Z} \hat{Z}$ \\
\hline
\end{tabular}
\caption{Coefficients and PS for the $(N_x, N_y) = (1,2)$ Ising Hamiltonian.}
\label{tab:ising_21}
\end{table}
\begin{table}[h!]
\footnotesize
\centering
\begin{tabular}{|c|c||c|c|}
\hline
\textbf{Coefficient $c_i$} & \textbf{PS $\hat{P}_{i}$} & \textbf{Coefficient $c_i$} & \textbf{PS $\hat{P}_{i}$} \\
\hline\hline
-0.005 & $\hat{X} \hat{I} \hat{I} \hat{I}$ & -0.071 & $\hat{I} \hat{X} \hat{I} \hat{Y}$ \\
\hline
-0.044 & $\hat{Y} \hat{I} \hat{I} \hat{I}$ & -0.096 & $\hat{I} \hat{X} \hat{I} \hat{Z}$ \\
\hline
1.028 & $\hat{Z} \hat{I} \hat{I} \hat{I}$ & -0.071 & $\hat{I} \hat{Y} \hat{I} \hat{X}$ \\
\hline
0.198 & $\hat{I} \hat{X} \hat{I} \hat{I}$ & -0.055 & $\hat{I} \hat{Y} \hat{I} \hat{Y}$ \\
\hline
-0.053 & $\hat{I} \hat{Y} \hat{I} \hat{I}$ & -0.039 & $\hat{I} \hat{Y} \hat{I} \hat{Z}$ \\
\hline
1.100 & $\hat{I} \hat{Z} \hat{I} \hat{I}$ & -0.096 & $\hat{I} \hat{Z} \hat{I} \hat{X}$ \\
\hline
-0.144 & $\hat{I} \hat{I} \hat{X} \hat{I}$ & -0.039 & $\hat{I} \hat{Z} \hat{I} \hat{Y}$ \\
\hline
0.161 & $\hat{I} \hat{I} \hat{Y} \hat{I}$ & 0.515 & $\hat{I} \hat{Z} \hat{I} \hat{Z}$ \\
\hline
0.893 & $\hat{I} \hat{I} \hat{Z} \hat{I}$ & -0.059 & $\hat{I} \hat{I} \hat{X} \hat{X}$ \\
\hline
-0.090 & $\hat{I} \hat{I} \hat{I} \hat{X}$ & -0.063 & $\hat{I} \hat{I} \hat{X} \hat{Y}$ \\
\hline
0.110 & $\hat{I} \hat{I} \hat{I} \hat{Y}$ & 0.101 & $\hat{I} \hat{I} \hat{X} \hat{Z}$ \\
\hline
0.805 & $\hat{I} \hat{I} \hat{I} \hat{Z}$ & -0.063 & $\hat{I} \hat{I} \hat{Y} \hat{X}$ \\
\hline
0.033 & $\hat{X} \hat{I} \hat{X} \hat{I}$ & -0.126 & $\hat{I} \hat{I} \hat{Y} \hat{Y}$ \\
\hline
0.014 & $\hat{X} \hat{I} \hat{Y} \hat{I}$ & 0.054 & $\hat{I} \hat{I} \hat{Y} \hat{Z}$ \\
\hline
0.067 & $\hat{X} \hat{I} \hat{Z} \hat{I}$ & 0.101 & $\hat{I} \hat{I} \hat{Z} \hat{X}$ \\
\hline
0.014 & $\hat{Y} \hat{I} \hat{X} \hat{I}$ & 0.054 & $\hat{I} \hat{I} \hat{Z} \hat{Y}$ \\
\hline
-0.028 & $\hat{Y} \hat{I} \hat{Y} \hat{I}$ & 0.645 & $\hat{I} \hat{I} \hat{Z} \hat{Z}$ \\
\hline
-0.162 & $\hat{Y} \hat{I} \hat{Z} \hat{I}$ & 0.067 & $\hat{Z} \hat{I} \hat{X} \hat{I}$ \\
\hline
0.413 & $\hat{Z} \hat{I} \hat{Z} \hat{I}$ & -0.149 & $\hat{X} \hat{X} \hat{I} \hat{I}$ \\
\hline
-0.154 & $\hat{X} \hat{Y} \hat{I} \hat{I}$ & -0.060 & $\hat{X} \hat{Z} \hat{I} \hat{I}$ \\
\hline
-0.154 & $\hat{Y} \hat{X} \hat{I} \hat{I}$ & -0.116 & $\hat{Y} \hat{Y} \hat{I} \hat{I}$ \\
\hline
0.027 & $\hat{Y} \hat{Z} \hat{I} \hat{I}$ & -0.060 & $\hat{Z} \hat{X} \hat{I} \hat{I}$ \\
\hline
0.027 & $\hat{Z} \hat{Y} \hat{I} \hat{I}$ & 0.429 & $\hat{Z} \hat{Z} \hat{I} \hat{I}$ \\
\hline
-0.076 & $\hat{I} \hat{X} \hat{I} \hat{X}$ & & \\
\hline
\end{tabular}
\caption{Coefficients and PS for the $(N_x, N_y) = (2,2)$ Ising Hamiltonian.}
\label{tab:ising_22}
\end{table}
\begin{table}[h!]
\footnotesize
\centering
\begin{tabular}{|c|c||c|c|}
\hline
\textbf{Coefficient $c_i$} & \textbf{PS $\hat{P}_{i}$} & \textbf{Coefficient $c_i$} & \textbf{PS $\hat{P}_{i}$} \\
\hline\hline
-0.005      & $\hat{X} \hat{I} \hat{I} \hat{I} \hat{I} \hat{I}$ &      0.136       & $\hat{I} \hat{Y} \hat{I} \hat{I} \hat{Z} \hat{I}$ \\
\hline
      -0.044      & $\hat{Y} \hat{I} \hat{I} \hat{I} \hat{I} \hat{I}$ &      0.004       & $\hat{I} \hat{Z} \hat{I} \hat{I} \hat{X} \hat{I}$ \\
      \hline
      1.028       & $\hat{Z} \hat{I} \hat{I} \hat{I} \hat{I} \hat{I}$ & 0.136       & $\hat{I} \hat{Z} \hat{I} \hat{I} \hat{Y} \hat{I}$ \\
      \hline
      0.198       & $\hat{I} \hat{X} \hat{I} \hat{I} \hat{I} \hat{I}$ & 0.559       & $\hat{I} \hat{Z} \hat{I} \hat{I} \hat{Z} \hat{I}$ \\
      \hline
      -0.053      & $\hat{I} \hat{Y} \hat{I} \hat{I} \hat{I} \hat{I}$ & 0.122      & $\hat{I} \hat{X} \hat{X} \hat{I} \hat{I} \hat{I}$ \\
      \hline
      1.100       & $\hat{I} \hat{Z} \hat{I} \hat{I} \hat{I} \hat{I}$ & 0.055       & $\hat{I} \hat{X} \hat{Y} \hat{I} \hat{I} \hat{I}$ \\
      \hline
      -0.144      & $\hat{I} \hat{I} \hat{X} \hat{I} \hat{I} \hat{I}$ & 0.156       & $\hat{I} \hat{X} \hat{Z} \hat{I} \hat{I} \hat{I}$ \\
      \hline
      0.161       & $\hat{I} \hat{I} \hat{Y} \hat{I} \hat{I} \hat{I}$ & 0.055       & $\hat{I} \hat{Y} \hat{X} \hat{I} \hat{I} \hat{I}$ \\
      \hline
      0.893       & $\hat{I} \hat{I} \hat{Z} \hat{I} \hat{I} \hat{I}$ & 0.127      & $\hat{I} \hat{Y} \hat{Y} \hat{I} \hat{I} \hat{I}$ \\
      \hline
      -0.090      & $\hat{I} \hat{I} \hat{I} \hat{X} \hat{I} \hat{I}$ & 0.087      & $\hat{I} \hat{Y} \hat{Z} \hat{I} \hat{I} \hat{I}$ \\
      \hline
      0.110       & $\hat{I} \hat{I} \hat{I} \hat{Y} \hat{I} \hat{I}$ & 0.156       & $\hat{I} \hat{Z} \hat{X} \hat{I} \hat{I} \hat{I}$ \\
      \hline
      0.805       & $\hat{I} \hat{I} \hat{I} \hat{Z} \hat{I} \hat{I}$ & 0.087      & $\hat{I} \hat{Z} \hat{Y} \hat{I} \hat{I} \hat{I}$ \\
      \hline
      0.061       & $\hat{I} \hat{I} \hat{I} \hat{I} \hat{X} \hat{I}$ & 0.584       & $\hat{I} \hat{Z} \hat{Z} \hat{I} \hat{I} \hat{I}$ \\
      \hline
      -0.007      & $\hat{I} \hat{I} \hat{I} \hat{I} \hat{Y} \hat{I}$ & 0.034      & $\hat{I} \hat{I} \hat{X} \hat{I} \hat{I} \hat{X}$ \\
      \hline
      0.950       & $\hat{I} \hat{I} \hat{I} \hat{I} \hat{Z} \hat{I}$ & 0.118      & $\hat{I} \hat{I} \hat{X} \hat{I} \hat{I} \hat{Y}$ \\
      \hline
      0.013       & $\hat{I} \hat{I} \hat{I} \hat{I} \hat{I} \hat{X}$ & 0.071      & $\hat{I} \hat{I} \hat{X} \hat{I} \hat{I} \hat{Z}$ \\
      \hline
      -0.264      & $\hat{I} \hat{I} \hat{I} \hat{I} \hat{I} \hat{Y}$ & 0.118      & $\hat{I} \hat{I} \hat{Y} \hat{I} \hat{I} \hat{X}$ \\
      \hline
      0.876       & $\hat{I} \hat{I} \hat{I} \hat{I} \hat{I} \hat{Z}$ & 0.001       & $\hat{I} \hat{I} \hat{Y} \hat{I} \hat{I} \hat{Y}$ \\
      \hline
      -0.076      & $\hat{X} \hat{I} \hat{I} \hat{X} \hat{I} \hat{I}$ &       0.12       & $\hat{I} \hat{I} \hat{Y} \hat{I} \hat{I} \hat{Z}$ \\
      \hline
      -0.071      & $\hat{X} \hat{I} \hat{I} \hat{Y} \hat{I} \hat{I}$ & 0.071      & $\hat{I} \hat{I} \hat{Z} \hat{I} \hat{I} \hat{X}$ \\
      \hline
      -0.096      & $\hat{X} \hat{I} \hat{I} \hat{Z} \hat{I} \hat{I}$ &       0.12       & $\hat{I} \hat{I} \hat{Z} \hat{I} \hat{I} \hat{Y}$ \\
      \hline
      -0.071      & $\hat{Y} \hat{I} \hat{I} \hat{X} \hat{I} \hat{I}$ & 0.539       & $\hat{I} \hat{I} \hat{Z} \hat{I} \hat{I} \hat{Z}$ \\
      \hline
      -0.055      & $\hat{Y} \hat{I} \hat{I} \hat{Y} \hat{I} \hat{I}$ & 0.078      & $\hat{I} \hat{I} \hat{I} \hat{X} \hat{X} \hat{I}$ \\
      \hline
      -0.039      & $\hat{Y} \hat{I} \hat{I} \hat{Z} \hat{I} \hat{I}$ & 0.06       & $\hat{I} \hat{I} \hat{I} \hat{X} \hat{Y} \hat{I}$ \\
      \hline
      -0.096      & $\hat{Z} \hat{I} \hat{I} \hat{X} \hat{I} \hat{I}$ & 0.068      & $\hat{I} \hat{I} \hat{I} \hat{X} \hat{Z} \hat{I}$ \\
      \hline
      -0.039      & $\hat{Z} \hat{I} \hat{I} \hat{Y} \hat{I} \hat{I}$ & 0.06       & $\hat{I} \hat{I} \hat{I} \hat{Y} \hat{X} \hat{I}$ \\
      \hline
      0.515       & $\hat{Z} \hat{I} \hat{I} \hat{Z} \hat{I} \hat{I}$ & 0.064       & $\hat{I} \hat{I} \hat{I} \hat{Y} \hat{Y} \hat{I}$ \\
      \hline
      -0.059      & $\hat{X} \hat{X} \hat{I} \hat{I} \hat{I} \hat{I}$ & 0.067      & $\hat{I} \hat{I} \hat{I} \hat{Y} \hat{Z} \hat{I}$ \\
      \hline
      -0.063      & $\hat{X} \hat{Y} \hat{I} \hat{I} \hat{I} \hat{I}$ & 0.068      & $\hat{I} \hat{I} \hat{I} \hat{Z} \hat{X} \hat{I}$ \\
      \hline
      0.101       & $\hat{X} \hat{Z} \hat{I} \hat{I} \hat{I} \hat{I}$ & 0.067      & $\hat{I} \hat{I} \hat{I} \hat{Z} \hat{Y} \hat{I}$ \\
      \hline
      -0.063      & $\hat{Y} \hat{X} \hat{I} \hat{I} \hat{I} \hat{I}$ & 0.437       & $\hat{I} \hat{I} \hat{I} \hat{Z} \hat{Z} \hat{I}$ \\
      \hline
      -0.126      & $\hat{Y} \hat{Y} \hat{I} \hat{I} \hat{I} \hat{I}$ & 0.061       & $\hat{I} \hat{I} \hat{I} \hat{X} \hat{I} \hat{X}$ \\
      \hline
      0.054       & $\hat{Y} \hat{Z} \hat{I} \hat{I} \hat{I} \hat{I}$ & 0.111      & $\hat{I} \hat{I} \hat{I} \hat{X} \hat{I} \hat{Y}$ \\
      \hline
      0.101       & $\hat{Z} \hat{X} \hat{I} \hat{I} \hat{I} \hat{I}$ & 0.076       & $\hat{I} \hat{I} \hat{I} \hat{X} \hat{I} \hat{Z}$ \\
      \hline
      0.054       & $\hat{Z} \hat{Y} \hat{I} \hat{I} \hat{I} \hat{I}$ & 0.111      & $\hat{I} \hat{I} \hat{I} \hat{Y} \hat{I} \hat{X}$ \\
      \hline
      0.645       & $\hat{Z} \hat{Z} \hat{I} \hat{I} \hat{I} \hat{I}$ & 0.169       & $\hat{I} \hat{I} \hat{I} \hat{Y} \hat{I} \hat{Y}$ \\
      \hline
      -0.003      & $\hat{X} \hat{I} \hat{X} \hat{I} \hat{I} \hat{I}$ &       0.03       & $\hat{I} \hat{I} \hat{I} \hat{Y} \hat{I} \hat{Z}$ \\
      \hline
      -0.001      & $\hat{X} \hat{I} \hat{Y} \hat{I} \hat{I} \hat{I}$ & 0.076       & $\hat{I} \hat{I} \hat{I} \hat{Z} \hat{I} \hat{X}$ \\
      \hline
      -0.121      & $\hat{X} \hat{I} \hat{Z} \hat{I} \hat{I} \hat{I}$ &       0.03       & $\hat{I} \hat{I} \hat{I} \hat{Z} \hat{I} \hat{Y}$ \\
      \hline
      -0.001      & $\hat{Y} \hat{I} \hat{X} \hat{I} \hat{I} \hat{I}$ & 0.484       & $\hat{I} \hat{I} \hat{I} \hat{Z} \hat{I} \hat{Z}$ \\
      \hline
      -0.059      & $\hat{Y} \hat{I} \hat{Y} \hat{I} \hat{I} \hat{I}$ & 0.152       & $\hat{I} \hat{I} \hat{I} \hat{I} \hat{X} \hat{X}$ \\
      \hline
      -0.042      & $\hat{Y} \hat{I} \hat{Z} \hat{I} \hat{I} \hat{I}$ & 0.038       & $\hat{I} \hat{I} \hat{I} \hat{I} \hat{X} \hat{Y}$ \\
      \hline
      -0.121      & $\hat{Z} \hat{I} \hat{X} \hat{I} \hat{I} \hat{I}$ & 0.011      & $\hat{I} \hat{I} \hat{I} \hat{I} \hat{X} \hat{Z}$ \\
      \hline
      -0.042      & $\hat{Z} \hat{I} \hat{Y} \hat{I} \hat{I} \hat{I}$ & 0.038       & $\hat{I} \hat{I} \hat{I} \hat{I} \hat{Y} \hat{X}$ \\
      \hline
      0.569       & $\hat{Z} \hat{I} \hat{Z} \hat{I} \hat{I} \hat{I}$ & 0.146      & $\hat{I} \hat{I} \hat{I} \hat{I} \hat{Y} \hat{Y}$ \\
      \hline
      0.094       & $\hat{I} \hat{X} \hat{I} \hat{I} \hat{X} \hat{I}$ & 0.012      & $\hat{I} \hat{I} \hat{I} \hat{I} \hat{Y} \hat{Z}$ \\
      \hline
      -0.144      & $\hat{I} \hat{X} \hat{I} \hat{I} \hat{Y} \hat{I}$ & 0.011      & $\hat{I} \hat{I} \hat{I} \hat{I} \hat{Z} \hat{X}$ \\
      \hline
      0.004       & $\hat{I} \hat{X} \hat{I} \hat{I} \hat{Z} \hat{I}$ & 0.012      & $\hat{I} \hat{I} \hat{I} \hat{I} \hat{Z} \hat{Y}$ \\
      \hline
      -0.144      & $\hat{I} \hat{Y} \hat{I} \hat{I} \hat{X} \hat{I}$ & 0.601       & $\hat{I} \hat{I} \hat{I} \hat{I} \hat{Z} \hat{Z}$ \\
      \hline
      -0.138      & $\hat{I} \hat{Y} \hat{I} \hat{I} \hat{Y} \hat{I}$ &                  &                               \\
\hline
\end{tabular}
\caption{Coefficients and PS for the $(N_x, N_y) = (2,3)$ Ising Hamiltonian.}
\label{tab:ising_23}
\end{table}

\end{document}